# NARROWBAND ACTIVE NOISE CONTROL FOR CONTINUOUS WAVE OPERATION OF TESLA CAVITIES

B. Richter*[1], J. Branlard, Deutsches Elektronen-Synchrotron DESY, Hamburg, Germany
A. Bellandi, Belgian Nuclear Research Centre SCK-CEN, Mol, Belgium
A. Benwell, J. Diaz-Cruz, S. Hoobler
SLAC National Accelerator Laboratory, Menlo Park, CA, USA
Q. Du, S. Murthy, Lawrence Berkeley National Laboratory, Berkeley, CA, USA
[1]also at Technische Universität Hamburg, Hamburg, Germany

*Abstract*

TESLA 9-cell elliptical cavities are used in the accelerators of the European XFEL and LCLS-II. The high duty cycle upgrade study for EuXFEL and the LCLS-II-HE program target loaded quality factors of $6 \times 10^7$ and beyond, leading to RF cavity half bandwidths below 11 Hz. Stable operation in these conditions heavily relies on active RF resonance control, since the RF drive power required to maintain amplitude and phase lock would surpass amplifier max ratings if detuning is not compensated. To counteract mechanical vibrations, e.g. introduced by rotational machines in the proximity of the accelerating modules, a dedicated narrowband active noise control algorithm was introduced in previous work. This proceeding presents CPU based implementations and latest test results obtained at the Cryomodule Test Bench (CMTB) at DESY and LCLS-II at SLAC, where the stable operation gradient of a cavity could be increased from 9 to 15 MV/m without violating RF control limits.

## INTRODUCTION

Latest XFELs all work towards CW operation of superconducting cavities for electron bunch acceleration. Energy consumption of superconducting cavity operation is mostly determined by required RF power to provide accelerating gradient and by cryogenic load through energy dissipation in cavity walls. The latter is determined by material properties and not changeable after manufacturing of the cavity. However, the required RF power $P_f$ to maintain an accelerating voltage $V_c$ depends on the cavity systems half bandwidth $f_{1/2}$ and detuning $\Delta f$

$$P_f = \frac{f_{1/2} V_c^2}{2 \frac{r}{Q} f_0} \left( 1 + \left[ \frac{\Delta f}{f_{1/2}} \right]^2 \right), \tag{1}$$

where beam loading is neglected [1]. Cavities are operated at $f_0 = 1.3$ GHz is a geometry specific constant. To minimize required amplifier peak power and operational power demand, the $f_{1/2}$ used with TESLA cavities operated in CW is generally chosen as low as possible, while still enabling to fulfill the required RF field stability criteria. However, with lower $f_{1/2}$, detuning effects on power requirement increase, therefore precise resonance control becomes crucial.

* bozo.richter@desy.de

Detuning arises through microscopic mechanical deformation of the cavity geometry. It originates from different sources: the accelerating RF field induces Lorenz force detuning through field interactions with the cavity walls; pressure variations of the liquid helium outside of the cavity leads to deformations; and spatial movements of cavity mounting points lead to deformation by inertia. Through these mounting points, rotational machinery like pumps and ventilators introduce vibrations into the cavity [2].

For active resonance control, TESLA cavities of XFEL, FLASH, and LCLS-II are equipped with piezo tuners allowing to apply unilateral longitudinal force to one of the cavity mounting points. We present practical benefits and implications of employing narrowband active noise control (NANC) to suppress periodic tune disturbances originating from rotational machinery.

## NARROWBAND ACTIVE NOISE CONTROL

The NANC algorithm is designed as an adaptive feedforward suppression controller for external harmonic disturbances. The adaption tracks slow disturbance drifts in amplitude, phase, and frequency, including gain and phase effects of the actuator and plant dynamics. It is not suitable to suppress mechanical resonances, which requires dedicated controller design. The notation used in this work is taken from Ref. [3], extended by the frequency adaption given in Ref. [4]. The algorithm is equivalent to that shown in Ref. [5], but different from the active resonance control filter bank of Ref. [6] previously implemented at LCLS-II.

NANC is based on harmonic drive signal components

$$u_m = \Re \left( A_m e^{i\omega_m t} \right), \tag{2}$$

defined by a drive vector $A_m \in \mathbb{C}$ and a drive frequency $\omega_m$. The effect of the actuation on the detuning is described by the detuning model

$$\Delta f_m = \Re \left( \tau_m A_m e^{i\omega_m t - i\Phi_m} \right), \tag{3}$$

additionally taking into account the forward gain $\tau_m$ and phase lag $\Phi_m$ of the drive path at the specific $\omega_m$. With this simple model, the measured detuning signal $\Delta f_{\text{meas}}$, and the external disturbance $\Delta f_{\text{mic}}$, the partial derivatives of a quadratic cost function

$$J_m = \Delta f_{\text{meas}}^2 = \left( \Delta f_{\text{mic}} + \sum_m \Delta f_m \right)^2, \tag{4}$$

can be used for drive parameter adaption using gradient descent, i.e.

$$A_m^+ = A_m - \mu \Delta f_{\text{meas}} A_m e^{i\omega_m t - i\Phi_m}, \tag{5}$$

$$\Phi_m^+ = \Phi_m - \eta \Delta f_{\text{meas}} \mathcal{I}\left(A_m e^{i\omega_m t - i\Phi_m}\right), \tag{6}$$

where the upper index + indicates the updated parameter, and $\tau_m$ and constant factors are absorbed into the learning rates $\eta$ and $\mu$. Note that when applied near resonances of the controlled system, both $\tau_m$ and $\Phi_m$ are subject to significant changes upon small variation of $\omega_m$, thus convergence will deteriorate, potentially even diverging.

If the external disturbance deviates from the defined $\omega_m$, the drive vector $A_m$ will exhibit a constant rotation to compensate the frequency mismatch. To offload the drive vector adaption and improve interpretability, the drive frequency $\omega_m$ can be adapted explicitly as presented in Ref. [4] by using the rotational variation of the drive vector correction, Eq. (5). By introducing the harmonic disturbance frequency $\omega_{\text{mic},m}$, the frequency error of a NANC component

$$e_{\omega,m} = \omega_{\text{mic},m} - \omega_m \tag{7}$$

$$= \nu\left[\arg\left(A_m^+\right) - \arg\left(A_m\right)\right], \tag{8}$$

with $\nu \in \mathbb{R}$ can be approximated for small deviations by

$$\begin{aligned} e_{\omega,m} &\approx \nu \frac{|A_m^+|}{|A_m|} \sin\left[\arg(A_m^+) - \arg(A_m)\right] \\ &= \nu \mathcal{I}\left(\frac{A_m A_m^+}{|A_m|^2}\right). \end{aligned} \tag{9}$$

The quadratic cost

$$J_{\omega,m} = e_{\omega,m}^2, \tag{10}$$

and the approximation Eq. (9) then lead to the gradient descent update

$$\omega_m^+ = \omega_m - \gamma \mathcal{I}\left(\frac{A_m A_m^+}{|A_m|^2}\right), \tag{11}$$

again absorbing constant factors into the learning rate $\gamma$. Notably, the frequency adaption is directly depending on availability of an updated drive vector $A_m^+$.

## IMPLEMENTATION AND RESULTS

The NANC algorithm is implemented in tests on CMTB at DESY and LCLS-II at SLAC. In both implementations, the detuning is calculated from RF data on external hardware, and then streamed to the resonance control hardware via low latency communication channels. On the resonance control FPGA, the detuning signal is decimated using a cascaded-integrator-comb filter and fed into a CPU, running a barebone application providing 4 harmonic NANC components per cavity. The main difference of this work and [4] is usage of an open source CPU architecture at LCLS-II. Details are given in Table 1. To ensure operational safety, saturation limits are implemented for adaptation of $A_m$, $\omega_m$, and $\Phi_m$.

| Facility | CMTB (DESY) | LCLS-II (SLAC) |
|---|---|---|
| Platform | DAMC-FMC25 | QF2-pre |
| FPGA | XC5VFX100T | XC7K160T |
| CPU | PPC SoC | RISCV [7] soft-core |
| NANC rate | 10 kHz | 5.5 kHz |

Table 1: General information of NANC implementations.

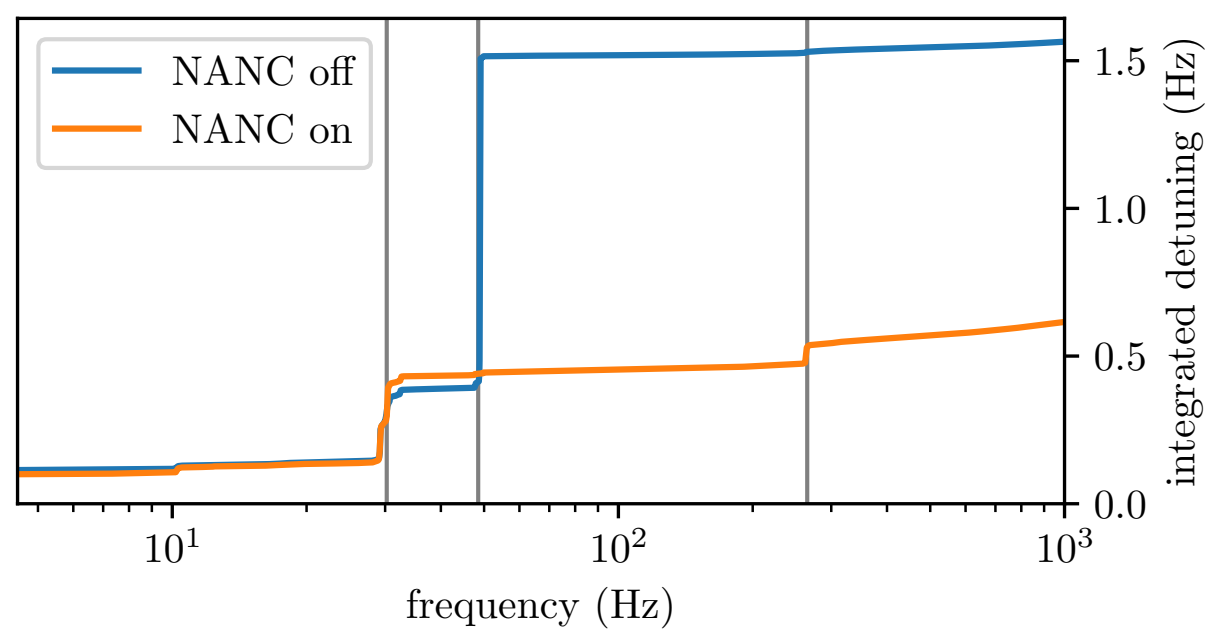

Figure 1: CMTB integrated detuning at 5 MV/m with and without a NANC component at 48 Hz. Final values mark detuning stds.

### CMTB

CMTB currently houses the spare cryomodule XM8 of the European XFEL, enabling R&D experiments for system enhancement and future upgrade studies. For the test of the NANC algorithm, cavity 5 is set to an external half bandwidth of 37.5 Hz and operated at a gradient of 5 MV/m. Detuning data is recorded for 120 s at 8 kS/s, first with NANC disabled, then with a single component enabled at 48 Hz.

The integrated power spectral density of the detuning is given in Fig. 1. NANC effectively cancels the 48 Hz disturbance and thereby reduces the detuning standard deviation (std) from 1.6 to 0.65 Hz, peaks from 6 to 4 Hz. However, disturbances at 30 and 265 Hz and wideband noise above 300 Hz increase in magnitude.

A second component placed at 30 Hz suffers from diverging $\omega_m$. Understanding of this divergence requires further investigation.

### LCLS-II

At LCLS-II, 280 TESLA cavities are operated at nominal $f_{1/2}$ of 16 Hz and accelerating gradients of 20 MV/m. The presented NANC controller is tested on two RF stations with cavities severely affected by narrow bandwidth tune disturbances. Here, we present results obtained at cavity 2 of cryomodule 19, where the gradient is limited to 9 MV/m due to microphonics.

First, the operating gradient is kept at 9 MV/m. Detuning data is recorded at 2 kS/s for 56 s, starting without NANC, then with activated NANC components at different frequencies. Frequencies are selected manually based on maximum peaks in detuning spectrograms. Histograms and integrated power spectral densities are shown in Fig. 2 and 3, respectively.

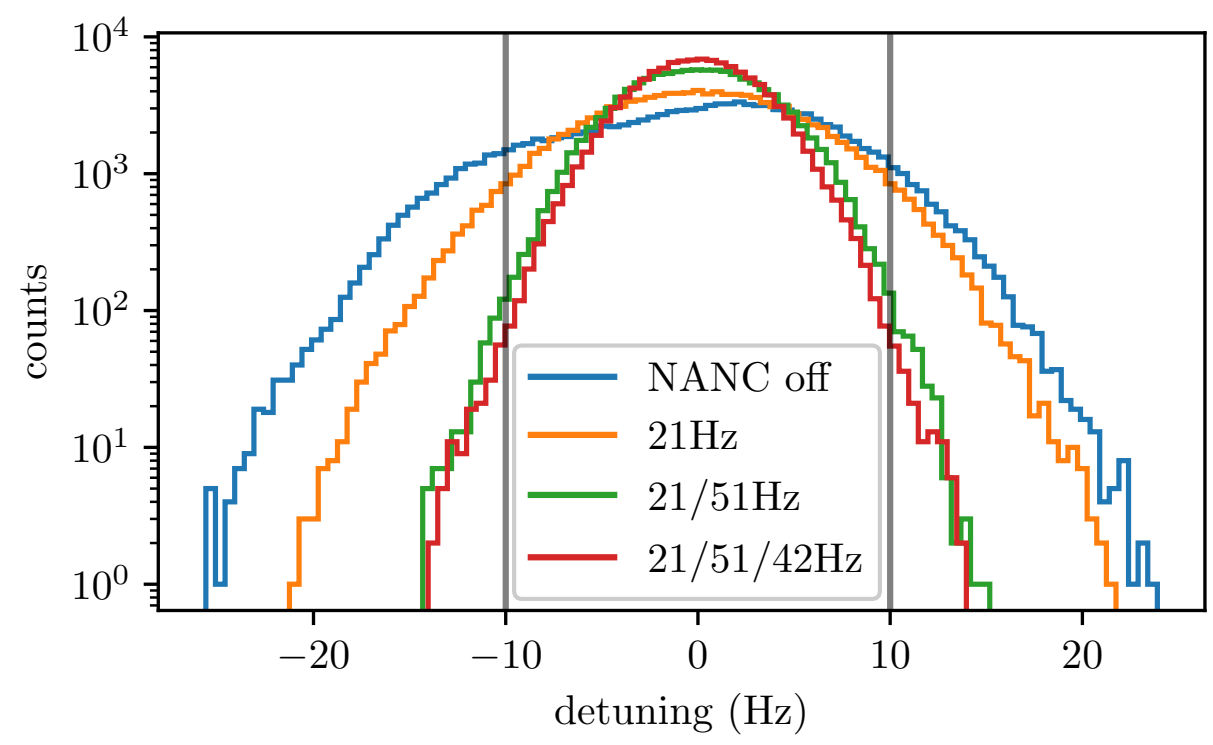

Figure 2: LCLS-II detuning histograms at 9 MV/m. NANC components are enabled sequentially at frequencies given in the legend, vertical lines mark specification limits.

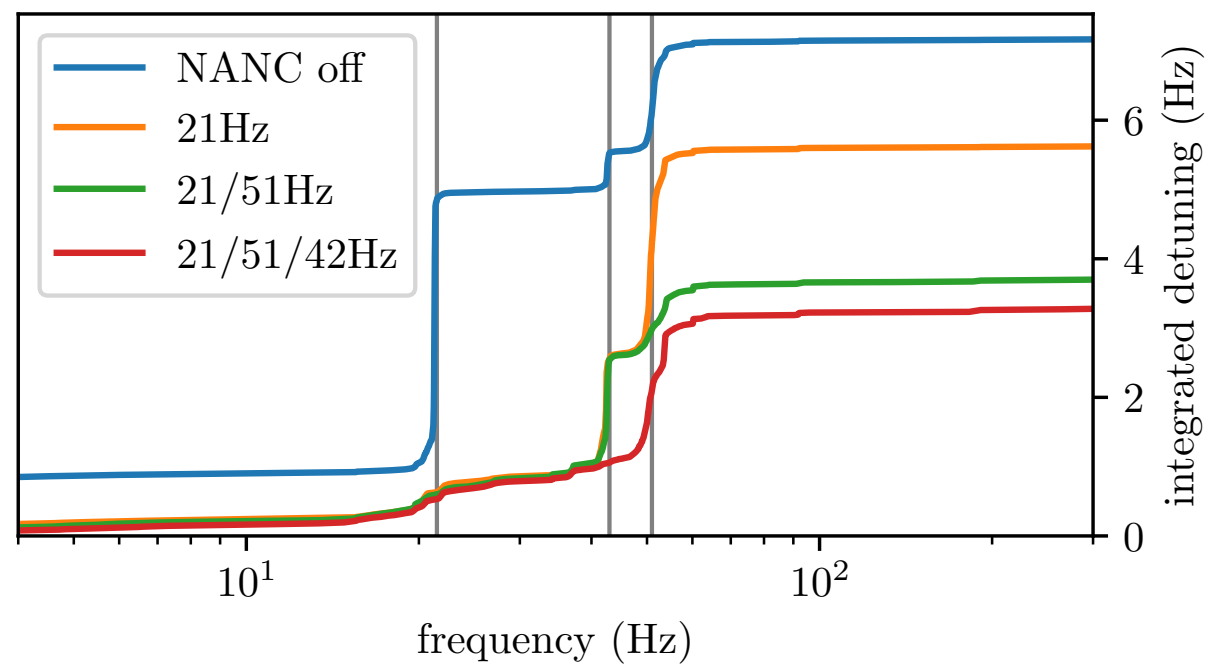

Figure 3: LCLS-II integrated detuning at 9 MV/m for subsequently enabled NANC components. Final values mark detuning std.

Figure 3 reveals that enabling the component at 21 Hz increases the disturbances at 42 Hz and 51 Hz, while considerably reducing those below 10 Hz. Subsequent activation of the second component at 51 Hz acts locally only without exciting other frequencies. Both the first and second component lead to a significant reduction of std and detuning peak values, see Fig. 2. Notably, around 50 Hz the disturbances are distributed over a band of a few Hertz, leaving room for improvement by wider band approaches.

The third component is placed at 42 Hz. It slightly reduces the detuning std, but barely shows an effect on peak detuning, as disturbances close to 51 Hz increase significantly, almost canceling the effects of disturbance reduction at 42 Hz.

Due to negligible effects, results of a fourth NANC component at 60 Hz are not shown. Overall, the detuning peak and std are reduced by a factor of two, leading to a significant reduction in required peak power. This reduction can either be leveraged by a more aggressive RF controller, translated to reduction in RF power consumption, or used to increase the acceleration gradient while maintaining stable operation. Nevertheless, the target specification for peak detuning of 10 Hz is not met.

A series of subsequent tests is performed by increasing the cavity gradient, where operation without NANC leads to occasional violation of RF power limits and severe performance degradation by loss of phase lock. Application of NANC allows increasing the gradient to 15 MV/m without violating these limits once over a monitoring timespan of more than 5 minutes. For analysis, detuning data is recorded at 2 kS/s for 160 s, first without NANC and then with the three components active as before. The histogram in Fig. 4 shows the occurrence of large detuning excursions originating from RF control limit violations (blue), and the stabilizing effect of NANC (red).

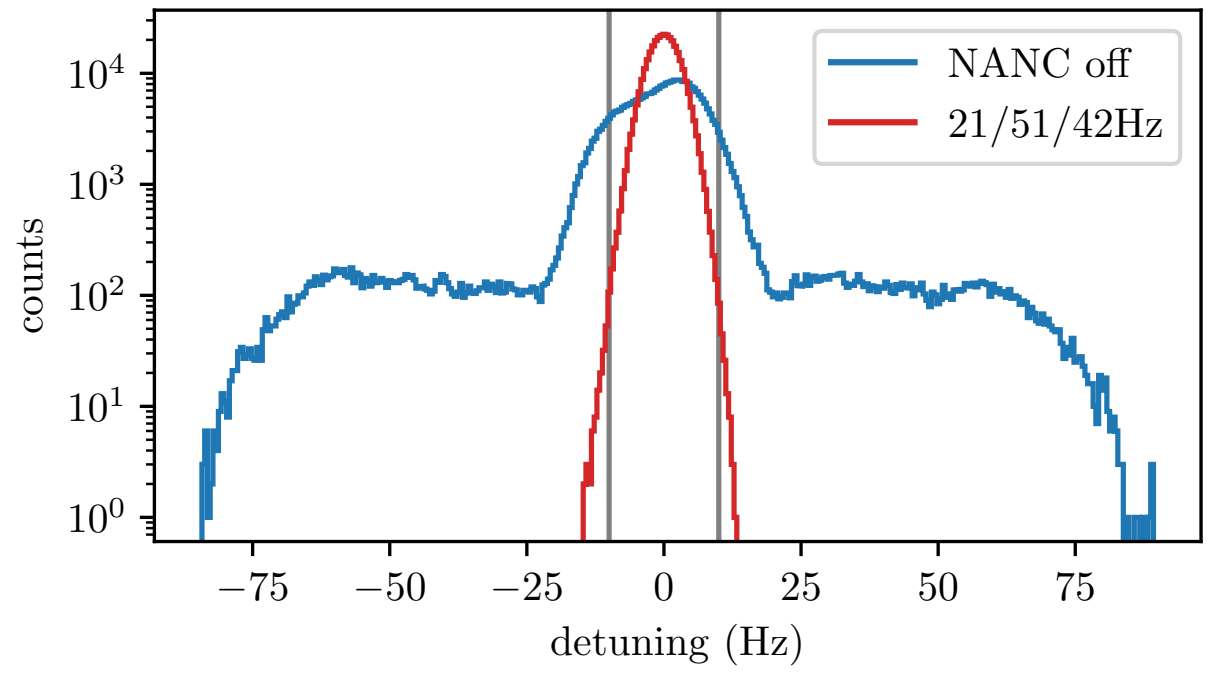

Figure 4: LCLS-II detuning histograms at 15 MV/m for NANC disabled and enabled.

## CONCLUSION AND OUTLOOK

The presented NANC approach is suitable for suppression of narrowband external disturbances sufficiently far from mechanical resonances, but not for damping of resonant modes. It offloads periodic disturbance compensation at specific frequencies to a dedicated adaptive feed-forward controller, complementing a feedback resonance controller.

Experimental results at CMTB and LCLS-II demonstrate that NANC effectively reduces detuning contributions from dominant harmonic disturbances. This leads to lower detuning standard deviation and peak values, and consequently reduced RF power demand. In particular, for cavities limited by microphonics, NANC enables stable operation at increased gradients without violating RF power constraints, while for well-performing cavities it provides potential for either improved energy efficiency or field stability.

Future work will address the applicability of NANC in pulsed operation in the context of the European XFEL high duty cycle upgrade.

## ACKNOWLEDGEMENTS

The authors acknowledge the scientific collaboration between SLAC and DESY that enabled the LCLS-II experiments presented in this work. This work was partially funded in the context of the R&D program of the European XFEL. This project has received funding from the European Unions Horizon Europe research and innovation program under grant agreement N. 101131435.